\documentclass[aip,prl,twocolumn,10pt,superscriptaddress,showkeys]{revtex4-1}
\usepackage{graphicx}
\usepackage{natbib}
\usepackage[section]{placeins}
\usepackage{setspace}

\begin{document}
\title{Dynamical Decoupling of a single electron spin at room temperature}
\author{Boris Naydenov$^{\mathrm{b)}}$}
\email{b.naydenov@physik.uni-stuttgart.de}
\affiliation{$3$ Physikalisches Institut and Research Center SCOPE, University of Stuttgart, Stuttgart 70659, Germany}
\author{Florian Dolde}
\thanks{These authors contributed equally}
\affiliation{$3$ Physikalisches Institut and Research Center SCOPE, University of Stuttgart, Stuttgart 70659, Germany}
\author{Liam T. Hall}
\affiliation{Centre for Quantum Computer Technology, School of Physics, University of Melbourne, Victoria 3010, Australia}
\author{Chang Shin}
\affiliation{National Biomedical Center for Advanced ESR Technology, Dept of Chemistry and Chemical Biology, Cornell University, Ithaca, NY 14853, USA}
\author{Helmut Fedder}
\affiliation{$3$ Physikalisches Institut and Research Center SCOPE, University of Stuttgart, Stuttgart 70659, Germany}
\author{Lloyd C.L. Hollenberg}
\affiliation{Centre for Quantum Computer Technology, School of Physics, University of Melbourne, Victoria 3010, Australia}
\author{Fedor Jelezko}
\affiliation{$3$ Physikalisches Institut and Research Center SCOPE, University of Stuttgart, Stuttgart 70659, Germany}
\author{J{\"o}rg Wrachtrup}
\affiliation{$3$ Physikalisches Institut and Research Center SCOPE, University of Stuttgart, Stuttgart 70659, Germany}

\bibliographystyle{apsrev4-1}
\begin{abstract}
Here we report the increase of the coherence time T$_2$ of a single electron spin at room temperature by using dynamical decoupling. We show that the Carr-Purcell-Meiboom-Gill (CPMG) pulse sequence can prolong the T$_2$ of a single Nitrogen-Vacancy center in diamond up to 2.44 ms compared to the Hahn echo measurement where T$_2 = 390~\mu$s. Moreover, by performing spin locking experiments we demonstrate that with CPMG the maximum possible $T_2$ is reached. On the other hand, we do not observe strong increase of the coherence time in nanodiamonds, possibly due to the short spin lattice relaxation time $T_1=100~\mu$s (compared to T$_1$ = 5.93 ms in bulk). An application for detecting low magnetic field is demonstrated, where we show that the sensitivity using the CPMG method is improved by about a factor of two compared to the Hahn echo method.
\end{abstract}
\pacs{61.72.jn,76.30.Mi,76.70.Hb,76.60.Lz}
\maketitle

\begin{figure}
\includegraphics[scale=0.28]{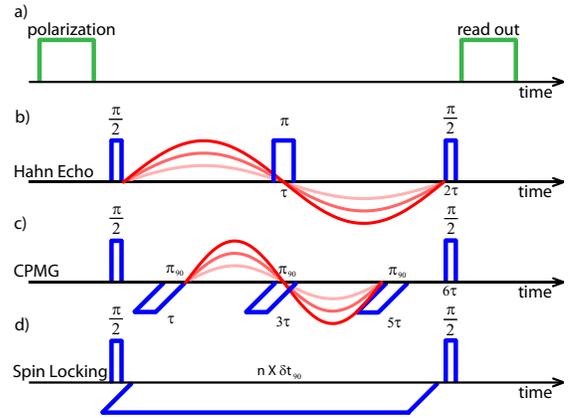} 
\caption{\label{PulseSeq}Pulse sequences used in the experiments. (a) Optical polarization and readout of the NV electron spin. (b) Hahn-echo. (c) CPMG. (d) Spin Locking (see text for more details).}
\end{figure}

\begin{figure}
\includegraphics[scale=0.5]{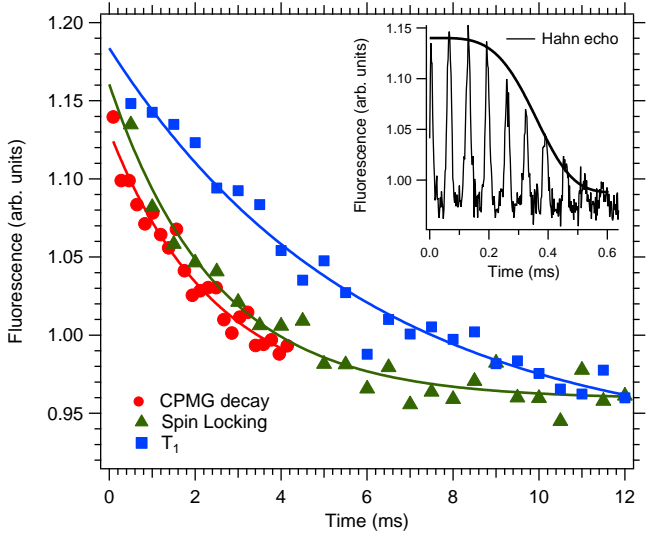} 
\caption{\label{HahnCPMG} Hahn echo decay (inset, $T_2=0.39\pm0.16$~ms), CPMG (red markers, $T_2^{\mathrm{CPMG}}=2.44\pm0.44$ms), spin locking (green markers, $T_{1\rho}=2.47\pm0.27$~ms) and spin lattice relaxation (blue, $T_1=5.93\pm0.7$~ms). The blue, red and green curves are fits to the data (see text).}
\end{figure}

Nitrogen-vacancy (NV) centers in diamond are one of the most promising quantum bits (qubits) for a scalable solid state quantum computer. Single NVs can be addressed optically even at room temperature \cite{Fedor04a, Fedor04b} and the first quantum registers containing several qubits have been demonstrated \cite{Dutt07,Philipp08,Philipp10}. One of the main advantages of the NV centers is their long coherence time T$_2$ at room temperature, reaching almost 2 ms in ultra pure isotopically enriched $^{12}$C diamond \cite{Gopi09}, permitting the detection of weak magnetic fields \cite{Maze08a, Gopi08}. The $T_2$ limited sensitivity was reported to be $4~\frac{nT}{\sqrt{Hz}}$, as measured in a $T_2=1.8$~ms sample \cite{Gopi09}. Recently, a wide field approach has been demonstrated  to have sensitivity of 20 $\frac{nT}{\sqrt{Hz}}$ where an ensemble of NVs are used as sensors \cite{Steinert10}.\\ 
It is of a crucial importance to develop new methods for increasing the coherence time of NV in a not ultra pure environment. It this Letter we demonstrate that $T_2$ of an NV center in a bulk diamond can be increased by a factor of six using the Carr-Purcell-Meiboom-Gill (CPMG) pulse sequence. The CPMG sequence is widely used in the NMR community \cite{CP56, MG58} and was recently rediscovered in the context of quantum computing theory  \cite{DasSarma07} and experiment \cite{Yacoby10,Liu09}, and has been proposed as a means to increase the sensitivity of NV based magnetometers \cite{Taylor08,Hall10}. Although common in the field of NMR, this sequence has not found wide application in electron spin resonance (ESR), as relatively few reports are known, for example \cite{Schmidt73, Eaton03}.\\
The NV center consists of a substitutional nitrogen atom and a neighboring  carbon vacancy. The system has triplet ground state as described by the following Hamiltonian ($\hbar=1$):
\begin{equation}
H=(\omega_L+\omega_e(t))S_z+D\left(S_z^2 + \frac{1}{3}S(S+1)\right)+H_{\mathrm{HF}}
\end{equation}
where $\omega_L=g_e\mu_BB_0$ is the Larmor frequency, $\omega_e(t)=g_e\mu_BB_e(t)$ represents the magnetic field fluctuations in the environment which cause decoherence, $g_e$ is the g-factor of the $S=1$ NV electron spin, $\mu_B$ is the Bohr magneton, $B_0$ is the applied constant magnetic field, $D=2.88$ GHz is the zero field splitting, $H_{\mathrm{HF}}$ is the hyperfine interaction to the nitrogen nucleus which may be ignored in the present context. A small magnetic field $B_0\approx15~$G is aligned along the NV quantization axis (defined by $D$, $z$ axis in in the rotating frame) in order to split the $m_s\pm1$ levels.
\begin{figure}[b]
\includegraphics[scale=0.6]{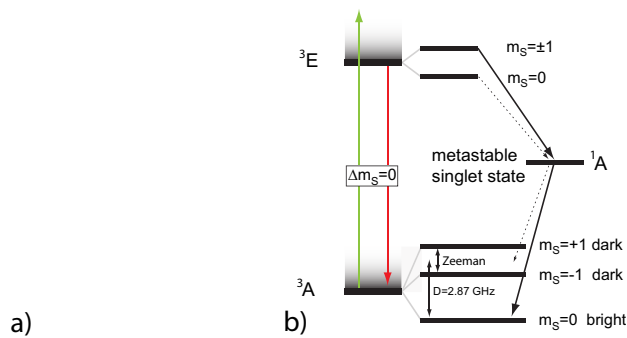} 
\caption{\label{NVScheme}(a) Schematic view of NV center in diamond. (b) Energy level scheme of NV. A green laser excites the NV to $^3E$, from it can fall back to $^3A$ or undergo inter system crossing to a meta stable state $^1A$. From there it decays to the $m_s=0$ ground state, thus polarizing the electron spin. }
\end{figure}
Aligning the field is important since $T_2$ depends on the orientation of $B_0$ where the maximum is reached for $B_0$ parallel to the NV axis \cite{Maze08,Stanwix10}.\\
The magnetic field at the NV center can written as $B(t)=B_0+B_e(t)$, where its mean field and its standard deviation are $\langle B\rangle=B_0$ and $B'(t)=\sqrt{\langle B^2\rangle-B_0^2}$. If the fluctuation rate of $B_e(t)$ is $f_e=1/\tau_e$ it has been recently shown \cite{Hall09} that in the case of slow fluctuation limit and for all $t$ ($1/\tau_eg_e\mu_BB'(t)1\gg1$) $B(t)$ can be expanded as a Taylor series:
\begin{equation}
B(t)=\sum_{k=0}^{N}\frac{1}{k}\frac{d^kB}{dt^k}\vert_{t_0}\equiv\sum_{k=0}^{N}a_k\left(t-t_0\right)^k\mathrm{,}
\label{FieldExpansion}
\end{equation}
where each $a_k$ represents a different dephasing channel.\\
We consider the Hahn echo pulse sequence depicted in Fig.~\ref{PulseSeq}b and its effect on eq.~\ref{FieldExpansion}. A laser pulse with wavelength $\lambda=532~$nm and 2 $\mu$s duration is used to polarize the NV into the $m_s=0$ state (Fig.~\ref{PulseSeq}a) and read out the population difference between $m_s=0$ and $m_s=1$ states (Fig.~\ref{NVScheme}b) \cite{Fedor04a}. A microwave (MW) $\pi/2$ pulse resonant with the $m_s=0\rightarrow m_s=+1$ transition, is applied along the $y$ axis in the rotating frame. The NV is transformed in to the superposition state $|\psi\rangle=1/\sqrt{2}(|0\rangle+|1\rangle)$ or equivalently the effect of the MW pulse is to transfer the equilibrium spin magnetization $M$ from $z$ to the $x$ axis in the rotating frame.
Inhomogeneities ($a_0$ in eq.~\ref{FieldExpansion}) and quasi-static fluctuations $B(t)$ due to the $^{13}$C spin bath in the surrounding of NV cause decay of $|\psi\rangle$ with decay function $e^{-t/T_2^*}$, process known as free induction decay (FID). The evolution of the system is described by the Hamiltonian $H_{\mathrm{evol}}=\omega(t)S_z$ and the evolution operator $U_{\mathrm{evol}}=e^{-\int{H_{\mathrm{evol}}}dt}$. The application of a $\pi$ at time $\tau$ inverts the sign of $H_{\mathrm{evol}}$, resulting in a "refocusing" of the spin coherence at time $2\tau$. Strictly speaking the contribution of $a_0$ in the dephasing is completely removed whereas the contribution from high order terms is suppressed via \cite{Hall09,Hall10}: 
\begin{equation}
a_k \mapsto \left(1-2^{-k}\right)a_k
\label{HahnRefocus}
\end{equation}
The final $\pi/2$ is used to transfer the coherence into population difference which is readout optically. With longer $\tau$, the effects on the phase coherence of low frequency fluctuations in the environment become more pronounced. Such effects can be mitigated significantly with the application of a CPMG pulse sequence \cite{DasSarma07} (Fig.~\ref{PulseSeq}c), in which a series of $\pi$ pulses are applied at times $2n+1$ for $n=0,1,...N$, yielding multiple echoes at times $2n+2$. The lower index in the rotation angle of the MW pulse represents the phase of the MW, where $0$ implies aligning the MW magnetic field $B_1\parallel y$ and $90$ $B_1\parallel x$. Note that the phase of the $\pi$ pulse train is shifted by $90^{\circ}$ ($B_1$ is aligned along $x$ of the rotating frame) in order to suppress errors in the pulse length \cite{MG58}.\\
The quantum phase accumulated by the NV spin, $\Delta\phi$, is proportional to the time integral of the magnetic field $B(t)$ from eq.~\ref{FieldExpansion}. If  $m$ pulses are applied at the instants $t_1,\,t_2,\,\ldots t_m$, the effect of the pulse sequence on the phase shift will be
\begin{eqnarray*}
  \Delta\phi &=& \gamma\left(\int_{0}^{t_1} - \int_{t_1}^{t_2} +\ldots+(-1)^m\int_{t_m}^\tau\right)B(t)\,dt.\label{effectonBt}
\end{eqnarray*}
The effect of an arbitrary sequence of pulses on the $j^\mathrm{th}$ term in the Taylor expansion is then
\begin{eqnarray*}
  a_j &\mapsto& a_j\frac{\left(\int_{0}^{t_1} - \int_{t_1}^{t_2}\ldots+(-1)^m\int_{t_m}^\tau\right)t^j\,dt}{\int_0^\tau t^j\,dt}.\label{effectonaj}
\end{eqnarray*}
For a CPMG sequence, the time of application of the $j$th pulse in an $n$ pulse sequence is $t_j =\frac{2j-1}{2n}$ where $j\in\{1,2,\ldots,n\}$. For 1 pulse (Hahn echo), the effect on the $k$th Taylor term is given in Eq.~\ref{HahnRefocus}. And in general for 2, 3 and n pulses, we have
\nonumber
\begin{eqnarray*}
  2\,a_k  &\mapsto& a_k\frac{1}{4^{k+1}} \left[2 - 2\left(3\right)^{k+1} + 4^{k+1}\right]\\
    3\,a_k  &\mapsto& a_k\frac{1}{6^{k+1}} \left[2 - 2\left(3\right)^{k+1} + 2\left(5\right)^{k+1} - 6^{k+1}\right]\\
    &\vdots&\nonumber\\
   n\,a_k &\mapsto& a_k\frac{1}{(2n)^{k+1}} [2 + (-1)^n(2n)^{k+1} +\\
    &+& 2\sum_{j=1}^{n-1}(-1)^j(2j+1)^{k+1} ]\\
\end{eqnarray*}
In the limit of $\tau\rightarrow0$ we arrive at the spin locking regime (Fig.~\ref{PulseSeq}d) where the system does not evolve freely, but it is constantly driven by the MW field. In this case the spin magnetization is "locked" to the $x$ axis in the rotating frame and it decays with time constant $T_{1\rho}$ determined by the noise spectral density $J(\omega_1)$ where $\omega_1=g\mu_{B_1}/\hbar$ is the Rabi frequency with $B_1$ the MW magnetic field. For very high MW power $\omega_1\approx\omega_L$ $T_{1\rho}$ approaches the spin-lattice relaxation time $T_1$ proportional to $J(\omega_L)$. Since the decoherence channels do not influence the transverse magnetization any more (the system does not evolve freely), $T_{1\rho}$ can be considered as the upper limit for $T_2$ measured by any multiple pulse (decoupling) sequence, if the same MW power is used.\\
The data from the coherence experiments are plotted in Fig.~\ref{HahnCPMG}.
The diamond sample used for these measurements has been CVD grown (Element 6) with natural $^{13}$C abundance (about 1\%) and low nitrogen impurity concentration (below 1 ppb). According to theoretical calculations the Hahn echo decays as $e^{-(\frac{2\tau}{T2})^4}$ \cite{deSousa09}, which was fitted to the data. From the fit we obtain $T_2=389~\mu$s, which is in very good agreement with the value predicted from the theory $T_{2}^{theory}=400~\mu$s for decoherence caused by fluctuations in the $^{13}$C ($I=1/2$, 1\% concentration) spin bath \cite{Hall10}. This electron-nuclear coupling results in the electron spin envelope modulation (ESEEM) at the Larmor frequency $\omega_C$ of $^{13}$C shown in Fig.~\ref{HahnCPMG}, as described in detail by Childress et al. \cite{Childress06}. For the CPMG measurement $\tau$ was set to be at the maximum of the Hahn echo revivals, thus providing maximum signal \cite{Eaton03}. If $\tau<2\pi/\omega_C$, we also observed oscillations in the echo train (data not shown). For the CPMG experiment the theory predicts exponential decay with increase of the decay constant as $T_2^{\mathrm{CPMG}}=(2n)T_2^{2/3}$, where $n$ is the number of pulses \cite{deSousa09}. We observe that $T_2^{\mathrm{CPMG}}=2.44$ ms, which about six time longer than $T_2$ whereas a factor of 32 ($n=90$) is expected from the theoretical formula. We also find out from the spin locking decay $T_{1\rho}=2.47$~ms, a value close to the measured spin lattice relaxation time $T_1=5.93$ ms. So we have $T_2^{\mathrm{CPMG}}=T_{1\rho}\sim T_1$, meaning that we are able to suppress the decoherence channels to the limit imposed by the relaxation processes $T_1$. Identical experiments war performed with nanodiamonds (ND, average diameter 30 nm, SYP), where the CPMG technique improves the $T_2$ only by a factor of two - from 2.1 $\mu$s to 4.8 $\mu$s. From spin locking measurements we extract $T_{1\rho}=13~\mu$s, suggesting that there is strong source of decoherence and relaxation in ND, which also limits $T_1$ to 100$\mu$s. This result could be explained by the large electron spin bath surrounding the NV in ND \cite{Tisler09}.\\ 
For the magnetometry experiments a gold microstructure was directly deposited on the diamond to provide MW and AC magnetic fields. The latter was created by an arbitrary waveform generator (Tektronix AWG 2041). The superposition state $|\psi\rangle$ during its free evolution accumulates a relative phase $\Delta \Phi$ which is used for the detection of small magnetic fields \cite{Taylor08,Gopi08}. The sensitivity is proportional to $\sqrt{T_2}$ and the collected phase is given by
\begin{equation}
\Delta \Phi = \int_0^\tau \Delta \omega dt = \frac{g_e\mu_B}{\hbar}\int_0^\tau B(t)dt
\label{phase}
\end{equation}
where $\Delta \omega$ is the shift of the Larmor frequency. The $\pi$ pulse in the Hahn echo changes the sign of the collected phase and $\Delta \Phi_{\mathrm{Hahn}}$ is then
\begin{equation}
\Delta \Phi_{\mathrm{Hahn}} = \frac{g_e\mu_B}{\hbar}\int_0^\tau B(t)dt-\frac{g_e\mu_B}{\hbar}\int_\tau^{2\tau} B(t)dt
\end{equation}
This measurement scheme can be used to detect AC magnetic field with frequency $\left(\frac{1}{2\tau}\right)$ and synchronized phase \cite{Maze08a}. For sensing with the CPMG pulse sequence, the frequency has to be set to $\frac{1}{4\tau}$ (see figure \ref{PulseSeq}). In this case $\Delta \Phi_{\mathrm{CPMG}}$ for n $\pi$ pulses is:
\begin{eqnarray*}
  \nonumber\Delta \Phi_{\mathrm{CPMG}} =\frac{g_e\mu_B}{\hbar}(\int_{0}^{\tau}B(t)dt\sum_{n=1,3,5,...}\int_{(2n-1)\tau}^{(2n+1)\tau}B(t)dt\\
  -\sum_{n=2,4,6,...}\int_{(2n-1)\tau}^{(2n+1)\tau}B(t)dt +(-1)^{n}\int_{(2n+1)\tau}^{(2n+2)\tau}B(t)dt  )
  \label{phaseCPMG}
\end{eqnarray*}
The detected signal is then proportional to $\cos(\Delta\Phi)$.
\begin{figure}
\includegraphics[width=0.5\textwidth]{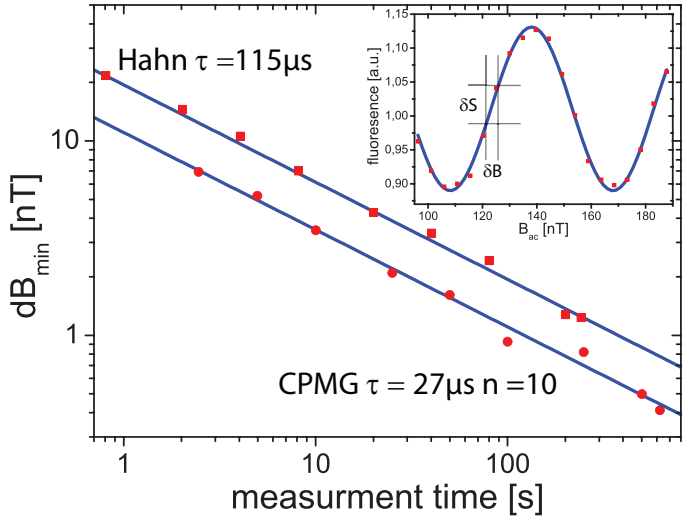}
\caption{The graph represents $\delta B_{min}$ as a function of the total measurement time per data point for Hahn echo and CPMG based magnetometery. The blue line are fits with the shot noise limit $\delta B_{min}=\frac{k}{\sqrt{t}}$. The inset shows the oscillations in the fluorescence intensity due to the applied AC magnetic field.}
\label{shotnoise}
\end{figure}
The lowest detectable magnetic field $\delta B_{min}$ is determined by the change of the measured signal and its error, where the steepest change in the signal is considered to maximize sensitivity. The error is given by the shot noise limitation of the collected photons. $\delta B_{min}$ can be calculated by
\begin{equation}\label{dBmin}
    \delta B_{min}=\frac{\sigma_{sn}}{\delta S}
\end{equation}
where $\sigma_{sn}$ is the uncertainty in the measured data point (determined by the standard deviation) and $\delta S$ is the steepness of the signal change (see the inset of Fig. \ref{shotnoise}). The dependence of $\delta B_{min}$ on $\sigma_{sn}$ is depicted in Fig. \ref{shotnoise}. $\tau=115~\mu$s was chosen for the Hahn echo based method. For the CPMG detection scheme $n=10$ and $\tau=27 \mu s$ was used. Increasing the number of pulses above ten did not improve the sensitivity, most likely due to fluctuations in the applied AC magnetic field. Nevertheless the application of the CPMG technique afforded significantly reduction of $\delta B_{min}$ as it can be seen in Fig.~\ref{shotnoise}. The fit of the shot noise limit $\delta B_{min}=\frac{k}{\sqrt{t}}$ yields a sensitivity $k$ of $k_{Hahn}=19.4\pm 0.4 \frac{nT}{\sqrt{Hz}}$ and $k_{CPMG}=11.0 \pm 0.2 \frac{nT}{\sqrt{Hz}}$.\\
We have demonstrated the possibility of extending the coherence times of a single NV centre in diamond via the application of a CPMG pulse sequence. Using this, we have managed to demonstrate improved AC magnetic field sensitivity compared to the Hahn echo method. These results open a new way towards the detection of single electron spins at ambient conditions which has wide application in life sciences and nanotechnology.\\
We are grateful to Gopalakrishnan Balasubramanian and Florian Rempp for useful discussions. This work is supported by the EU (QAP, EQUIND, NEDQIT, SOLID), DFG (SFB/TR21 and FOR730, FOR1482), BMBF (EPHQUAM and KEPHOSI) and Landesstiftung BW. LCLH and LTH acknowledge support of the Australian Research Council.\\
Results closely related to this work have been recently reported by de Lange et al. \cite{Lange10} and Ryan et al. \cite{Cory10}.

\end{document}